\documentclass[aip,amssymb,reprint,nofootinbib]{revtex4-1}

\usepackage{amsmath}
\usepackage{hyperref}

\usepackage{graphicx}
\usepackage{dcolumn}
\usepackage{bm}

\usepackage[utf8]{inputenc}
\usepackage[T1]{fontenc}
\usepackage{mathptmx}\usepackage{etoolbox}
\usepackage{xspace}

\newcommand{\fket}[1]{|#1\rangle}
\newcommand{\ket}[1]{|#1\rangle}
\newcommand{\PYTHEUS}{{\texttt{PYTHEUS}}\xspace}

\begin{document}

\title{Terry vs an AI, Round 1: Heralding single-rail (approximate?) 4-GHZ state from squeezed sources.} 

\author{Terry Rudolph}
\affiliation{Dept. of Physics, Imperial College London, London SW7 2AZ}
\affiliation{PsiQuantum, Palo Alto.}
\email{t.rudolph@imperial.ac.uk}
\date{\today}

\begin{abstract}
The potential for artificial intelligence (AI) to take over the work of physicists should be treated with glee. Here I evaluate one of the scientific discoveries in quantum photonics made by a leading AI in the field, in order to try and gain insight into when I will be allowed to go spend my days sipping mezcal margaritas on a warm beach. My analysis leads me to the distressing conclusion that it may, in fact, be quite a while yet. 
\end{abstract}

\pacs{}

\maketitle 


\emph{This paper is prepared for submission to the Jonathan P. Dowling memorial issue of AVS Quantum Science. I have tried to write it in a polemical but hopefully still fun style that Jon would have appreciated. Any implied criticisms in this paper are only directed at non-human members of the respective collaborations. I am the last person who wants to discourage work in theoretical linear optics by any sort of intelligence(s)!}

\section{Introduction}

Recently the world was introduced to \PYTHEUS \cite{PYTHEUS1,PYTHEUS2}. Like his homophonic namesake, \PYTHEUS is an explorer, who in just a few short years has already made new scientific discoveries\cite{PYTHEUS2}!

\PYTHEUS is an artificial intelligence who cannot (yet) embark on physical voyages, and so his discoveries at present focus on finding linear-optical circuits (interferometers) that take squeezed-light inputs and optimize for creating targeted output photonic states. Now this is a task that I personally am pretty good at. And frankly, while half the world is justifiably panicked about losing their jobs to an AI, I would be absolutely delighted to have this aspect of my research usurped. 

Moreover, superhuman chess-playing AI's have not destroyed the game. On the contrary, humans enjoy learning new strategies from them. So part of me wondered if I could learn something from \PYTHEUS. But another part of me is just competitive - is \PYTHEUS really the Deep Blue of quantum optics? Am I a Kasparaov? Carlsen? Perhaps together \PYTHEUS and I could at least be a Niemann?

I decided to pick one example of a discovery made by \PYTHEUS, and see if I could re-discover it for myself. But before turning to my concrete attempt at such, let me overview the basis for some of my skepticism that I will be able to hand off my job to an AI any time soon. 

 I strongly suspect \PYTHEUS is relying on the ingenuity of his human collaborators far more than he lets on. Here are a few examples of things in which I have participated that I consider at least mildly interesting ``scientific discoveries'' of the last couple of years. They all involve only understanding scattering of photons through interferometers (\PYTHEUS' claimed area of expertise) but I do not believe \PYTHEUS has any chance (in his current state of mind) of emulating them without considerable assistance from the humans he is shackled to:
\begin{itemize}
\item From four single photons a dual-rail Bell state can be created with probability $2/3$, which is significantly greater than $1/4$, the presumed upper bound confirmed by more than a decade's worth of numerical searches (see Section V of Ref.~\onlinecite{ssgpaper} - therein termed ``bleeding'', could there be any stronger way of rubbing it in?!?).
\item Bell state measurement vs Bell basis measurement? A useful linear-optical POVM that projects onto dual-rail Bell states can be performed with higher probability than a projective Bell basis measurement. (Section VIIC of Ref.~\onlinecite{ssgpaper}). 
\item It is possible, with only polynomial growth in the number of modes, to do linear-optical quantum computing using no coherent switch at all, only a ``blocking'' (i.e. absorbing/incoherent) switch\cite{tezmultirail}. Moreover this can be done using only probabilisitic single photon sources. This shows multiple core assumptions about what is necessary for photonic quantum computing are simply false. 
\item One can deterministically create photonic entanglement which mimics the structure of first quantization using second quantized states\cite{aliens} (``third quantization''), and this enables serious things like hiding into thermal light correlations an alien civilization's distributed quantum computations, as well as mad possibilities like simulating indistinguishable particles that can nevertheless be individually accessed, or performing purely W-state driven measurement-based quantum computing. 
\end{itemize}
 
Each of those examples requires one to think outside of standard paradigms, despite being only simple photonic interferometry.  The first requires realizing there can be immense power in protocols that have intermediate adapativity and do not terminate in a fixed depth of interferometer. The second requires understanding that projection onto non-orthogonal Bell states is just as useful for almost everything in quantum information as projection into a (partial) Bell basis - and then finding a better-performing, ancilla assisted effective POVM for such a measurement. The third requires realizing that coherent erasure of information can be arranged to underpin all linear optical quantum computing. The fourth requires...? I guess it requires you to be somewhat crazy and desperate to push the ultimate limits of single photon entanglement to try and understand something that photons cannot do, only to realize that nope, they actually can do it! I suspect \PYTHEUS could discover some technical aspects of all the above if well guided, but not off his own bat. 

A final reason I feel I'm not going to be out of the race with \PYTHEUS any time soon is that simplicity is crucial for practical, large-scale photonic engineering. Unless care is taken to reduce experimental complexity, it is easy to end up needing $O(n^2)$ components to implement an $n$ mode experiment, and such should typically be avoided if possible. Seems clear that considerations of ease and robustness will help us poor humans remain in the race a little longer - the best experiments are the ones as dumb as their designers!

\section{A concrete challenge}

Many of the experiments \PYTHEUS has been thinking about are for postselected state generation. Personally I find this very uninteresting, because (i) modern quantum technologies need, at a minimum, heralded state generation and (ii) the Reeh-Schlieder theorem already tells me I can postselect an elephant out of the vacuum with a suitable measurement. So as a challenge I decided to focus on \PYTHEUS's solution for the heralded generation of a superposition of vacuum and four single photons - i.e. single-rail (approximate) 4-GHZ state - given in Fig.~4 of Ref.~\onlinecite{PYTHEUS2}.

\begin{figure}
 \includegraphics[width=8.5cm]{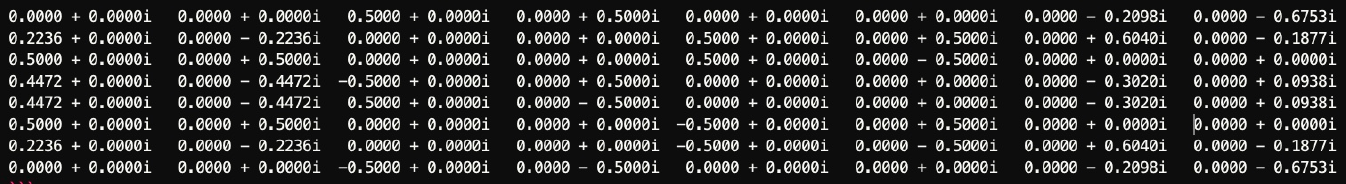}
 \caption{\label{fig:PYTHEUSunitary} Unitary matrix for an 8-port interferometer which \PYTHEUS discovered can herald the generation of the 4-photon, single-rail, approximate 4-GHZ state.}
 \end{figure}

To actually convert the graph of Fig.~4(b) to a standard interferometric setup I enlisted a (purportedly human) assistant Jake Bulmer, who reported back\footnote{\href{https://github.com/jakeffbulmer/terry\_vs\_ai/blob/main/jakes\_translations\_for\_terry\_vs\_AI.ipynb}{\texttt{https://github.com/jakeffbulmer/terry\_vs\_ai/blob/main/ jakes\_translations\_for\_terry\_vs\_AI.ipynb}}} that \PYTHEUS claims all we need to do is take 6 single mode squeezed state sources, choosing the specific values for the squeezing parameters of: $r = [0.9350,  0.9350,  0.7849,  0.7849, 0.3403, 0.3403]$ along with two vacuum mode ancillae, and feed them all through the interferometer of Fig.~\ref{fig:PYTHEUSunitary}. After this, heralding single photons on modes $3,4,5,6$ will create the (Fock basis) state $|0000\rangle+\sqrt{\epsilon} |1111\rangle$  in modes $1,2,7,8$, where $\epsilon\approx 1\%$. The probability of the scheme succeeding is also about $1\%$. By changing a scaling factor in the solution we can increase $\epsilon$ but the success probability goes down even further. 

Problematically, Jake noticed (and I confirmed with Matlab) that it appears there are also small amplitude terms in the output state of the form $|2110\rangle$ and $|0112\rangle$. Are these real, or just an artifact of poor numerical precision??

As you can see we have already encountered a problem many humans are having in their conversations with famous AI's like AlphaZero, DALL-E and so on: they are very good at popping out magical and impressively complicated looking results, but do so without providing any clear explanation as to what is actually going on in their machinations.\footnote{Those of us used to doing standard dumb numerical optimizations, for which impregnable numerical output is of course the norm, are almost tempted to claim that perhaps these AI's are convincing their human friends to totally exaggerate how smart and earth-shattering they are. But I digress.}  

Now before going further let me point out that even if correct, the usefulness of this specific problem for real-world challenges such as building a photonic quantum computer is debatable. Firstly, making direct use of states with coherence between different photon numbers is problematic, mainly because loss is the main error in quantum optics and it becomes no longer heraldable. Dephasing between the subspaces of different total photon number is often also an issue. Secondly, heralding entanglement generation at such low success probabilities means that to use it fruitfully the whole setup will need to be \emph{multiplexed}, i.e. repeated many (hundreds) of times  -  spatially or temporally - and then very large suitable optical switches somehow employed to select a success.

We can greatly ameliorate this latter issue by breaking any heralded state generation protocol into as small a set of heraldable chunks as possible, and multiplexing those separately/sequentially (e.g. see section VI of  Ref.~\onlinecite{ssgpaper} for an example, therein termed ``primates'', aptly named in homage of its creators!). But if we are only given a numerical solution of the form ``complicated states in, big interferometer, big mess out, detect'' then determining how to do any such intermediate multiplexing itself becomes a completely new challenge.  

In the following I set out to both confirm and understand \PYTHEUS's discovery of his ``new actionable concept in physics'', hoping to break it down to pieces which are more conducive to the shallow-learning that meat-based intelligences are capable of. As such I try and explain my thought processes, including erroneous steps I made.

\section{First thoughts}

Jake's decomposition of \PYTHEUS's solution uses single mode squeezing. But the squeezing parameters come in pairs, and doing a balanced beamsplitter on two identical single-mode squeezed states produces a two-mode squeezed state (TMSS) of the form
\begin{equation}
|S\rangle=s_0|00\rangle+s_1|11\rangle+s_2|22\rangle+s_3|33\rangle+\ldots
\end{equation}
 Thus one might hope to instead start with three TMSSs which is conceptually easier for me. Even simpler would be if we only had to interfere the idler modes. But there is no way that taking any number of TMSSs, interfering only the idlers then detecting them all can create a coherent superposition between vacuum and any other number state (the same total photon number measured on the idlers must output on the signals).

\begin{figure}
 \includegraphics[width=6.5cm]{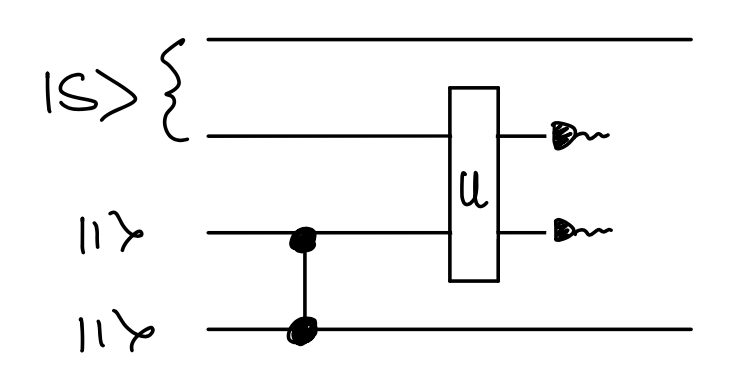}
 \caption{\label{fig:int1} Interferometric circuit for which detection of 2 photons yields an output that has coherence only between the vacuum and 4-photon subspaces. $U$ is arbitrary. Note this is not a qubit circuit, here lines are modes connected via interferometers. The vertical line with black dots is a 50:50 beamsplitter which creates $\fket{20}-\fket{02}$ from the $\fket{11}$ part of the  input. See Ref.~ \onlinecite{ssgpaper} for more on this notation.}
 \end{figure}

Hmm, what about interfering signals and idlers independently, then detecting some of the signals and some of the idlers? This is more optimistic, but it does seem like a strong restriction (only being able to interfere the signals with each other and the idlers with each other). Ok, well perhaps I can step back and just convince myself of one simpler thing: \emph{
Conjecture: to create a superposition between a vacuum and any state of 4 photons, with no terms containing $1$ or $2$ or $3$ or $\ge5$  photons, one must begin with at least three TMSS's.} I smacked my head on proving just this for a while, it is so obviously true, but to no avail.

\section{Tez's idiocy revealed}

One of my primary stumbling points was the difference between trying to make ``a superposition of vacuum and four photons, plus some (error) terms of more than four photons'' and ``a superposition containing \emph{only} terms of vacuum and four photons''. The former is much easier - for example if you take a single photon and a TMSS you can use the HOM effect to strip out the $|11\rangle$ term from the squeezed state. But you will still have higher order errors ($|33\rangle$ terms etc.)

At some point I suddenly realized I'm being an idiot. Take a general TMSS input into an interferometric circuit like Fig.~\ref{fig:int1}. If you detect exactly two photons then either they both came from the bunched single photons initially in modes 3,4, or they both came from the TMSS. Either way, none of the terms $\fket{nn}$ of $\ket{S}$, with $n=1,3,4,\ldots$ can herald the detection pattern, and so we do have a state with \emph{exactly} vacuum and four photons (though not the desired $\fket{1111}$ state).

\begin{figure}
 \includegraphics[width=6.5cm]{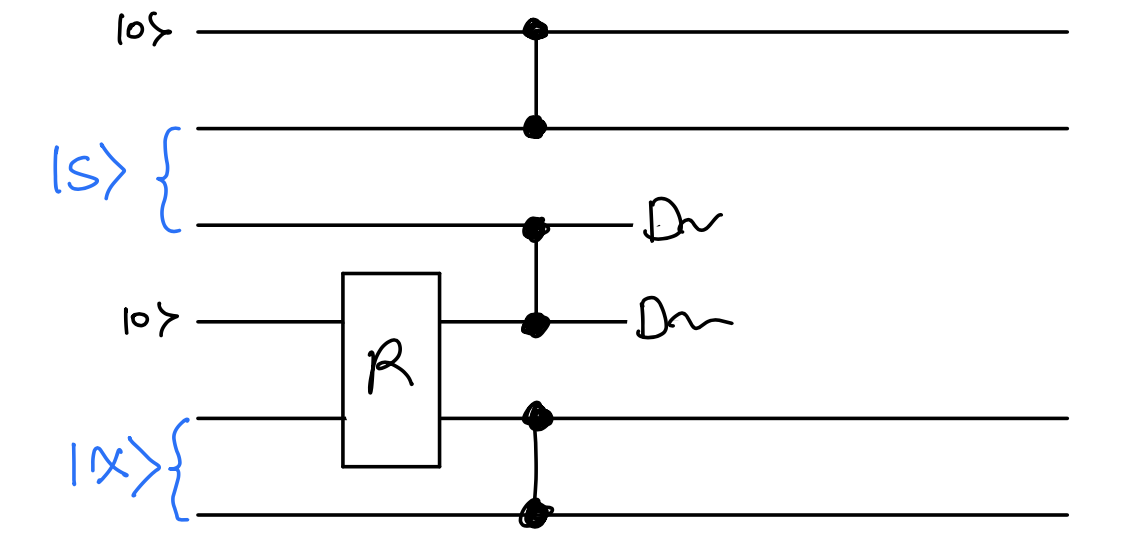}
 \caption{\label{fig:int2} Interferometric circuit for which detection of 2 photons yields an output that has coherence only between the vacuum and 4-photon subspaces, but now with 4-mode output that is much closer to the target state. $R$, $\ket{\chi}$ are given in the text.}
 \end{figure}

From here we can readily get to the slightly more complicated circuit of Fig.~\ref{fig:int2} where $R=\begin{bmatrix} \sqrt{a} & \sqrt{1-a} \\ \sqrt{1-a} &-\sqrt{a}\end{bmatrix}$ and $\ket{\chi}=\sqrt{b} \fket{20}-\sqrt{1-b}\fket{02}.$  

Consider we herald on detecting two single photons. Again the output is a superposition of either vacuum or 4 photons. A calculation simple enough even for Maple (a program nobody would mistake for intelligent) tells us that the amplitude for vacuum at the output is
$$x_{0000}=-\tfrac{s_0}{\sqrt{2}}(1-a)\sqrt{b}.$$ Defining $\beta_{\pm}=-s_2(\sqrt{1-b}\pm a\sqrt{b})/4$ Maple says the amplitude for output $\fket{1111}$ is
$$x_{1111}=\sqrt{2}\beta_+.$$ Moreover the only other nonzero amplitudes are:
$$
     -x_{1     1     0     2}=
     -x_{1     1     2     0}=
     x_{0     2     0     2}=
     x_{0     2     2     0}=
     x_{2     0     0     2}=
     x_{2     0     2     0}=\beta_-
$$ 
and 
$$
     x_{0     2     1     1}=
     x_{2     0     1     1}=-\beta_+.
$$

We see that by choosing $\beta_-=0$, i.e. taking $a^2=b^{-1}-1$ (and $b\in[1/2,1]$) we interfere away the amplitudes of $\fket{1120}$ and $\fket{1102}$ as well as a bunch of other undesirable looking terms. It seems we end up with unavoidable amplitude in $\fket{0211}$ and $\fket{2011}$. These are the exact same error terms as Matlab claims for \PYTHEUS's solution. Hmm.

The success probability is 
$$ P_{succ}=s_0^2(\tfrac{1}{2}-\sqrt{b(1-b)})+s_2^2(1-b).$$ Choosing optimal squeezing for single photon generation ($s_0^2=1/2$, $s_1^2=1/4$, $s_2^2=1/8$) we can find a value of $b$ for which $\epsilon=x_{1111}^2/P_{succ}= 1\%$ to match \PYTHEUS.  However my $P_{succ}$ will be much higher, because my circuit makes use of a ``pre-given'' input state $\ket{\chi}$ which is a potentially unbalanced superposition of $\fket{20}$ and $\fket{02}$. This reiterates a point from earlier - this smaller initial state could be created offline and then multiplexed.   

How could $\ket{\chi}$ be created? The obvious method is from two TMSS's create single photons, use a 50:50 beamsplitter to create the balanced superposition $\fket{20}+\fket{02}$ and then probabilistically ``damp'' one term by coupling one mode to a vacuum ancilla with a beamsplitter of suitable transmittivity and herald on vacuum in the ancilla. Less obvious is to use two TMSS's and interfere the heralds with vacuum ancillae modes, before selecting 2 photon events. Interestingly, for $b>1/2$ this latter procedure is more efficient than the obvious one.

Note that regardless of how I produce $\ket{\chi}$ I will be using 3 TMSS's total, which is equivalent to \PYTHEUS's method in terms of initial Gaussian resources. But the $\sqrt{2}$ factor difference between the amplitudes of $x_{1111}$ and the error term amplitudes $x_{0211}$, $x_{2011}$ shows that by adding in a suitable 50:50 beamsplitter the output state of my solution will actually be of the form 
\begin{equation}
\ket{\Psi_{tez}}=\sqrt{c}\ket{0000}+\sqrt{1-c}\ket{2110}.
\end{equation}
This leads me to admit:
\begin{quotation}
{\noindent If it is true that \PYTHEUS's solution is not equivalent (via linear-optical interferometer) to a state of this form then he wins the match!} 
\end{quotation}

I suspect, but am not sure, my solution not only incorporates \PYTHEUS's as a special case, it is also objectively superior because mine makes clear both how to herald states with a wider variety of amplitudes and success probabilities, as well as to see the options for intermediate multiplexing\footnote{Look \PYTHEUS, if you \emph{have} made an error then be a bit of a (human) gentleman about it and simply insist such was your intention all along. You can justify why we'd be interested in this state as follows: Imagine we have the ability to easily create states of the form $\fket{000}+\sqrt{\epsilon}\fket{k11}$ (alt. $\fket{00}+\sqrt{\epsilon}\fket{k2}$) for any $k\ge 1$. Then from $N\approx 1/\epsilon$ copies it is possible with probability $>1/e$ to herald creation of a state which is convertible (via linear optics) into a maximally entangled qu\emph{d}it state, in a nice and robust $d$-rail encoding, with $d=N$ (alt. $d=N/2$). This is a lot of entanglement, created with high probability! Can you see how? If so please apply to interns@psiquantum.com.}. I would like to say the jury is still out, but then I should probably channel Jon Dowling:  \begin{quotation}{\noindent NO MATTER WHAT YOU SAY I DECLARE MYSELF THE ONLY ALLOWED CONTESTANT, THE SOLE JUDGE AND THE SUPREME WINNER OF THE MATCH!}\end{quotation}

[Aside: The truth is I can immediately think of many ways of heralding, from Gaussian sources, a superposition comprising strictly only vacuum or the state $|1111\rangle$. For example, with two copies of $|\Psi_{tez}\rangle$ you could simply put the first mode from each on a 50:50 beamsplitter and herald on detecting 2 photons. But its all much easier once you understand that using a TMSS in a quantum scissors protocol \cite{scissors} allows one to asymmetrically chop off terms with $>1$ photons in any mode while simultaneously doing the (Fock basis) bit flip $\ket{0} \leftrightarrow \ket{1}$. Note, however, that these schemes would all use more than the $6$ single-mode squeezed states of \PYTHEUS's claimed solution.]

 \section{Musings}

For reasons stated earlier I am not necessarily interested in the particular example considered here per se. But I am very interested in whether \PYTHEUS or his friends can help move the needle in the quest to build a quantum computer. It seems to me the main way this could happen is for us to teach \PYTHEUS about the primary sources of error in photonic quantum computing (imperfect multimode sources, losses, manufacturing imperfections in passive interferometers and so on), and then ask him to think about the most robust methods of creating the (highly constrained) types of photonic entanglement which can actually be used for fault tolerant computation.


\bibliography{LOQCpapers5.bib}

\end{document}